%% file: AggregDens2012-07-17Arxiv.tex
\documentclass[12pt]{article}
\usepackage{amsmath}
\usepackage{amsfonts}
\usepackage[english]{babel}
\usepackage{graphicx}

\begin{document}

\title{Aggregating density estimators: an empirical study}

\author{M. Bourel\footnote{M. Bourel, mbourel@fing.edu.uy, IMERL, Facultad de Ingenieria, Universidad de la Rep\'ublica, Julio Herrera y Reissig 565, 11300 Montevideo, Uruguay.}, B. Ghattas\footnote{B. Ghattas, badih.ghattas@univ-amu.fr, Universit\'e d'Aix Marseille, Institut de Math\'ematiques de Luminy, case901, 163 avenue de Luminy,  marseille 13009, France}}

\maketitle

\begin{abstract}
We present some new density estimation algorithms obtained by bootstrap aggregation like Bagging. Our algorithms are analyzed and empirically compared to other methods found in the statistical literature, like stacking and boosting for density estimation. We show by extensive simulations that ensemble learning are effective for density estimation like for classification. Although our algorithms do not always outperform other methods, some of them are as simple as bagging, more intuitive and has computational lower cost.
\end{abstract}


\maketitle

\section{Introduction}
Ensemble learning is one of the most challenging recent approaches in statistical learning. Bagging (\cite{Brei-bag}), Boosting (\cite{Freund}), Stacking (\cite{Brei-stack}), and Random forests (\cite{RF}) have been declared to be the best of the chelf classifiers achieving very high performances when tested over tens of various datasets selected from the machine learning benchmark. All these algorithms had been designed for  supervised learning, sometimes initially restricted to regression or binary classification. Several extensions are still under study: multivariate regression, multiclass learning, and adaptation to functional data or time series. \\
Very few developments exist for ensemble learning for unsupervised framework, clustering analysis and density estimation. Our work concerns the latter case which may be seen as a fundamental problem in statistics. Among the last developments, we found some extensions of boosting (\cite{Freund}) and  stacking (\cite{Smyth}) to density estimation. \\
In this paper we suggest some simple algorithms for density estimation in the same spirit of bagging and stacking where the weak learners are histograms. We show by extensive simulations that aggregation gives rise to effective better estimates. We compare our algorithms to several algorithms for density estimation, some of them are simple like Histogram and kernel density estimators (KDE) and others rather complex like stacking and boosting which will be described in details. As we will show in the experiments we do, although the accuracy of our algorithm is not systematically higher than other ensemble methods, it is with no doubt simpler, more intuitive and computationally less expensive.
Boosting algorithms and stacking for density estimation are described in section 2. Section 3 describes our algorithms. Simulations and results are given in section 4 and concluding remarks and future work in section 5.

\section{A review of the existing algorithms}

In this section we review some density estimators obtained by aggregation. They may be classified in two categories depending on the aggregation form.\\
The first type has the form of linear or convex combination:
\begin{equation}
f_M(x)= \sum_{m=1}^M \alpha_m g_m(x)
\label{LC}
\end{equation}

where $g_m$ is typically a parametric or non parametric density model, and in general different values of $m$ refer typically to
\begin{itemize}
\item different parameters values in the parametric case or,
\item different kernels, or
\item different bandwidths for a chosen kernel for the kernel density estimators.
\end{itemize}
The second type of aggregation is multiplicative and is based on the ideas of high order bias reduction for kernel density estimation (\cite{Jones}). The aggregated density estimator has the form:
\begin{equation}
f_M(x)= \prod_{m=1}^M \alpha_m g_m(x)
\label{BR}
\end{equation}

\subsection{Linear or convex combination of density estimators}
This kind of estimators (\ref{LC}) has been used in several works with different construction schemes.
\begin{itemize}
\item In \cite{Rosset}, \cite{Ridgeway} and \cite{Song} the weak learners $g_m$ are introduced sequentially in the combination. At step $m$, $g_m$ is chosen to maximize the log likelihood of
\begin{equation} f_{m}(x)= (1 - \alpha) f_{m-1}(x) + \alpha g_m(x) \label{maj} \end{equation}
where $g_m$ is a density selected among a fixed class $\mathcal{H}$.\\
In \cite{Rosset} $g_m$ is selected among a non parametric family of estimators, and in \cite{Ridgeway} and \cite{Song}, it is taken to be a Gaussian density or a mixture of Gaussian densities whose parameters are estimated. Different methods are used to estimate both density $g_m$ and the mixture coefficient $\alpha$.
In \cite{Ridgeway} $g_m$ is a Gaussian density and the log likelihood of (\ref{maj}) is maximized using a special version of Expectation Maximization (EM) taking into account that a part of the mixture is known.

The main idea underlying the algorithms given by \cite{Rosset} and \cite{Song} is to use Taylor expansion around the negative log likelihood that we wish to minimize:

$$ \sum_i - \log (f_{m}(x_i)) = \sum_i - \log (f_{m-1}(x_i)) - \alpha \sum_i \frac{g_m(x_i)}{f_{m-1}(x_i)} + O(\alpha^2) $$
 For $\alpha$ small we have the approximation

$$ \sum_i - \log (f_{m}(x_i)) \sim \sum_i - \log (f_{m-1}(x_i)) - \alpha \sum_i \frac{g_m(x_i)}{f_{m-1}(x_i)} $$

thus, minimizing the left side term is equivalent to maximizing $\sum_i \frac{g_m(x_i)}{f_{m-1}(x_i)}$.

All the algorithms described above are sequential and the number of weak learners aggregated may be fixed by the user.

\item \cite{Smyth} use stacked density estimator applying the same aggregation scheme as in stacked regression and classification (\cite{Wolpert}). The $M$ densities $g_m$ are fixed in advance (KDE with different bandwidths). The data set $\mathcal{L}=\{x_1,\dots,x_n\}$ is divided into $V$ cross validation subsets $\mathcal{L}_1,\dots,\mathcal{L}_V$. For $v=1,..,V$, denote $\mathcal{L}^{(-v)}=\mathcal{L}-\mathcal{L}_v$.
The $M$ models $g_1,\dots,g_M$ are fitted using the training samples $\mathcal{L}^{(-1)},\dots,\mathcal{L}^{(-V)}$, the obtained estimates are denoted $\widehat{g}_m^{(-1)},\dots,\widehat{g}_m^{(-V)}$ for all $m=1,\dots,M$. These models are then evaluated over the test sets $\mathcal{L}_1,\dots,\mathcal{L}_{V}$, getting the vectors $\widehat{g}_m^{(-v)}(\mathcal{L}_v)$ for $m=1,\dots,M,\, v=1,\dots,V$ put within a $n \times M$ matrix

$$
A=\left(
\begin{array}{cccc} \widehat{g}_1^{(-1)}(\mathcal{L}_1) & \dots & \dots & \widehat{g}_M^{(-1)}(\mathcal{L}_1) \\
    \widehat{g}_1^{(-2)}(\mathcal{L}_2) & \dots & \dots & \widehat{g}_M^{(-2)}(\mathcal{L}_2)\\
    \vdots & & & \vdots \\
    \widehat{g}_1^{(-V)}(\mathcal{L}_{V}) & \dots & \dots & \widehat{g}_M^{(-V)}(\mathcal{L}_{V})
\end{array}\right)
$$

This matrix is used to compute the coefficients $\alpha_1, \dots, \alpha_M$ using the Expectation-Maximization algorithm. Finally, for the output model, we re-estimate the individual densities $g_1,\dots,g_M$ from the whole data $\mathcal{L}$.

\item In \cite{Rigo} the densities $\{g_m\}_{m=1,...,M}$ are fixed in advance like for stacking (KDE estimators with different bandwidths). The dataset is split in two parts. The first sample is used to estimate the densities $g_m$, whereas the coefficients $\alpha_m$ are optimized using the second sample. The splitting process is repeated and the aggregated estimators for each data split are averaged. The final model has the form
$$ f_M(x) = \frac{1}{card\{S\}}\sum_{s \in S} \tilde{g}^s_M(x)$$
where $S$ is the set of all the splits used and
$$ \tilde{g}^s_M(x) = \sum_{m=1}^M \alpha_m g_m^s(x)$$
is the aggregated estimator obtained from one split $s$ of the data, $g_m^s$ is the individual kernel density function estimated over the learning sample obtained from the split $s$. This algorithm is called {\it AggPure}.
\end{itemize}

\subsection{Multiplicative aggregation}
\begin{figure}[h]
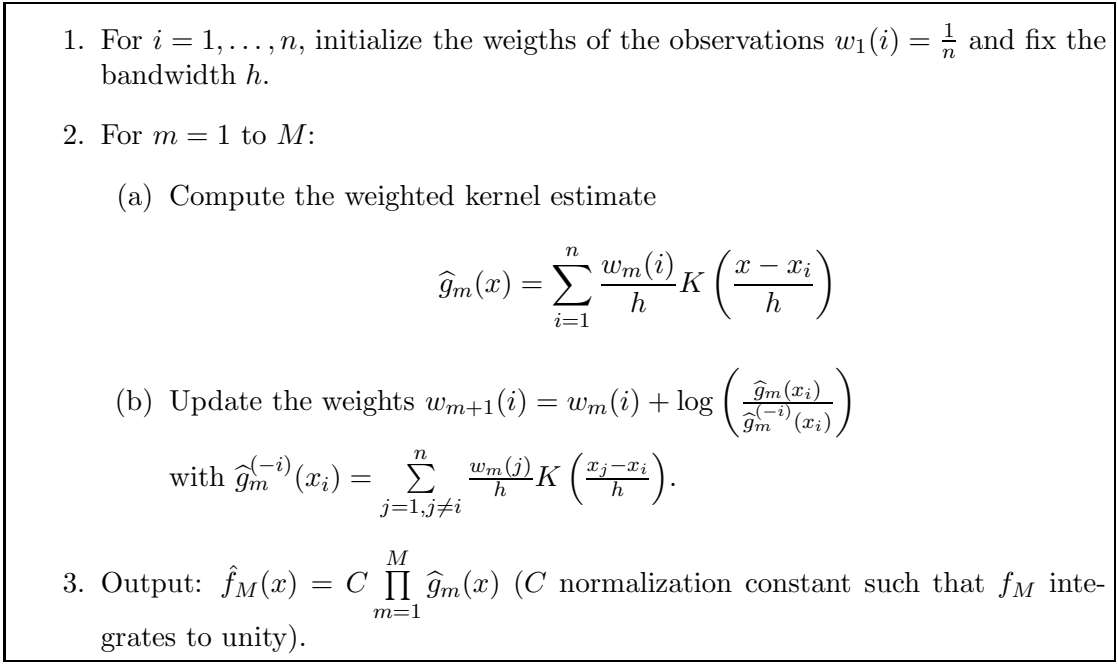

\fbox{
\begin{minipage}{0.9\textwidth}
{\small \begin{enumerate}
\item For $i=1,\dots,n$, initialize the weigths of the observations $w_1(i)=\frac{1}{n}$ and fix the bandwidth $h$.
\item For $m=1$ to $M$:
\begin{enumerate}
\item Compute the weighted kernel estimate $$\widehat{g}_m(x)= \sum \limits_{i=1}^{n} \frac{w_m(i)}{h}K \left(\frac{x-x_i}{h} \right)$$
\item Update the weights $w_{m+1}(i)=w_{m}(i) + \log \left(\frac{\widehat{g}_m(x_i)}{\widehat{g}_{m}^{(-i)}(x_i)} \right)$

with $\widehat{g}_{m}^{(-i)}(x_i)=\sum \limits_{j=1,j \neq i}^n \frac{w_m(j)}{h}K\left(\frac{x_j - x_i}{h}\right)$.
\end{enumerate}
\item Output: $\hat{f}_M(x)= C \prod \limits_{m=1}^{M}\widehat{g}_m(x)$ ($C$ normalization constant such that $f_M$ integrates to unity).
\end{enumerate} }
\end{minipage}}
\caption{Boosting kernel density estimation algorithm ((BoostKde), \cite{DiMarz}) \label{Dimarz}}
\end{figure}

The only algorithm giving rise to this form of aggregation is the one described in \cite{DiMarz} called {\it BoostKde}. It is a sequential algorith where at each step $m$ the weak learner is computed as follows:
$$\hat{g}_m(x)=\sum \limits_{i=1}^{n} \frac{w_m(i)}{h}K\left(\frac{x-x_i}{h}\right)$$
where $K$ is a fixed kernel, $h$ its bandwidth, and $w_m(i)$ the weight of observation $i$ at step $m$. Like for boosting, the weight of each observation is updated:
$$w_{m+1}(i)=w_m(i)+\log\left(\frac{\hat{g}_m(x_i)}{\hat{g}_{m}^{(-i)}(x_i)}\right)$$

where $\hat{g}_{m}^{(-i)}(x_i)=\sum \limits_{j=1,j \neq i}^n \frac{w_m(j)}{h}K\left(\frac{x_j - x_i}{h}\right)$.
The output is given by $\hat{f}_M(x)= C \prod \limits_{m=1}^{M}\hat{g}_m(x)$, where $C$ is a normalization constant. The Algorithm is resumed in figure \ref{Dimarz}.

\section{Aggregating Histograms}

We suggest three new density estimators obtained by linear combination like in (\ref{LC}), all of them use histograms as weak learners. The first two algorithms may be parallelized and randomize the histograms. The third one is just a modification of Stacking.

The first algorithm is similar to Bagging (\cite{Brei-bag}). At each step $m$, a bootstrap sample of the original dataset is generated and $g_m$ is the histogram obtained from the generated bootstrap sample with a fixed number of equally spaced breakpoints. We will refer to this algorithm as {\it BagHist}, it is detailed in figure \ref{Baghist}.

\begin{figure}[!ht]
\fbox{
\begin{minipage}{0.9\textwidth}
\hspace{5mm}
{\small \begin{enumerate}
\item Let $\mathcal{L}$ be the original sample
\item For $m=1$ to $M$:
\begin{enumerate}
\item Let $\mathcal{L}^{m}$ be a bootstrap sample of $\mathcal{L}$
\item Set $g_m$ to be the histogram constructed over $\mathcal{L}^m$ with equispaced $L$ breakpoints.
\end{enumerate}
\item Output: $f_M(x)= \frac{1}{M} \sum_1^M g_m(x)$ \\
\end{enumerate} }
\end{minipage}}
\caption{Bagging of histograms ({\it BagHist})}
\label{Baghist}
\end{figure}

The second algorithm ({\it AggregHist}) works as follows: let $g_0$ be the histogram obtained with the data set at hand using equally spaced breakpoints denoted $\mathcal{B} = {b_1,b_2,...,b_L}$. Each weak learner $g_m$ is an histogram constructed over the same initial data set but using a randomly modified set of breakpoints; $\gamma$ is a tuning parameter which controls the variance of the perturbations. The algorithm is detailed in figure \ref{Aggreghist}.

\begin{figure}[!ht]
\fbox{
\begin{minipage}{0.9\textwidth}
\hspace{5mm}
{\small \begin{enumerate}
\item Let $\mathcal{L}$ be the original sample, $g_0$ be the histogram constructed over $\mathcal{L}$ and \\ $\mathcal{B} = \{b_1,b_2,...,b_L\}$ the set of the ordered optimized breakpoints.
\item For $m=1$ to $M$:
\begin{enumerate}

\item Set $\mathcal{B}^m = \{b^*_{(1)},b^*_{(2)},...,b^*_{(L)}\}$ the modified breakpoints obtained by setting \\ $b^*_l =b_l + \varepsilon_l$ where $\varepsilon_l \sim N(0,\sigma)$ and $\sigma = \gamma \; min_{1 < l \le L} \left\{ b_{l} - b_{l-1} \right\}$.
\item Set $g_m$ to be the histogram constructed over $\mathcal{L}$ using the breakpoints $\mathcal{B}^m$.
\end{enumerate}
\item Output: $f_M(x)= \frac{1}{M} \sum_1^M g_m(x)$ \\
\end{enumerate} }
\end{minipage}}
\caption{Aggregating histograms based on randomly perturbed breakpoints ({\it AggregHist})}
\label{Aggreghist}
\end{figure}

Finally, we introduce a third algorithm called {\em StackHist} where we replace in the stacking algorithm described in the previous section, the six kernel density estimators by three histograms with fixed number of breaks.

The values of the parameters used in these algorithms will be optimized. The procedure used will be described in the experiments section.
\section{Experiments}
We test several simulation models based on classical distributions and mixture models mostly used in the cited works and algorithms described above. The sample size is fixed at $n=100, 500, 1000$. \\
We first show the estimates obtained using {\it BagHist} and {\it AggregHist} and analyze the effect of the number $M$ of histograms aggregated, and then compare them to the other algorithms. \\
\indent Up to our knowledge the existing algorithms for density estimation by aggregation have never been compared over a common benchmark simulation data.

\subsection{Models used for the simulations}

We denote by $\mathcal{M}1,\dots,\mathcal{M}11$ the different simulation models which will be grouped according to their difficulty level.

\begin{itemize}
\item Some standard densities used in \cite{DiMarz}:

    ($\mathcal{M}1$): standard Gaussian density $N(0,1)$

    ($\mathcal{M}2$): standard exponential density $f(x)=\left\{\begin{array}{lc} 0 & x<0\\ e^{-x}& x \geq 0\end{array} \right.$

    ($\mathcal{M}3$): a Chisquare density $\chi^2_{10}$

    ($\mathcal{M}4$): a Student density $t_4$

\item Some Gaussian mixtures taken from \cite{DiMarz} and \cite{Smyth}:

    ($\mathcal{M}5$): $0.5 N(-1,0.3)+0.5N(1,0.3)$

    ($\mathcal{M}6$): $0.5 N(-2.5,1)+0.5N(2.5,1)$

    ($\mathcal{M}7$): $0.25N(-3,0.5)+0.5N(0,1)+0.25N(3,0.5)$

\item Gaussian mixtures used in  from \cite{Rigo} and taken from \cite{Marron}

    ($\mathcal{M}8$): the Claw density, $0.5 N(0,1)+\sum \limits_{i=0}^{4}\frac{1}{10}N\left(\frac{i}{2}-1,\frac{1}{10}\right)$

    ($\mathcal{M}9$): the Smooth Comb Density,
$\sum \limits_{i=0}^{5} \frac{2^{5-i}}{63} N \left(\frac{65-96\frac{1}{2^{i}}}{21}, \frac{\left(\frac{32}{63}\right)^2}{2^{2i}}\right)=$

\scriptsize
    \begin{equation*}
    \frac{32}{63}N \left(-\frac{31}{21}, \frac{32}{63}\right)+\frac{16}{63}N \left(\frac{17}{21}, \frac{16}{63}\right)+\frac{8}{63}N \left(\frac{41}{21}, \frac{8}{63}\right)+\frac{4}{63}N \left(\frac{53}{21}, \frac{4}{63}\right)+\frac{2}{63}N \left(\frac{59}{21}, \frac{2}{63}\right)+\frac{1}{63}N \left(\frac{62}{21}, \frac{1}{63}\right)
    \end{equation*}
\normalsize
\item Mixtures density with highly inhomogeneous smoothness as in \cite{Rigo}:

    ($\mathcal{M}10$): $0.5 N(0,1) + 0.5 \sum \limits_{i=1}^{10} \mathbf{1}_{\left(\frac{2(i-1)}{T},\frac{2i-1}{T}  \right]}$

    ($\mathcal{M}11$): $0.5 N(0,1) + 0.5 \sum \limits_{i=1}^{14} \mathbf{1}_{\left(\frac{2(i-1)}{T},\frac{2i-1}{T}  \right]}$

\end{itemize}

All the simulations are done with the R software, and for models $\mathcal{M}8$ and $\mathcal{M}9$ we use the {\sf benchden} package.

We show below in figures \ref{dens1}, \ref{dens2} and \ref{dens3}, the true densities for the eleven models as well as their estimates obtained using the three algorithms {\it AggregHist}, {\it BagHist} and {\it StackHist} for $n=500$ observations and $M=300$ histograms for the two first algorithms.

\begin{figure}[!ht]
\centering
\includegraphics[scale=0.31]{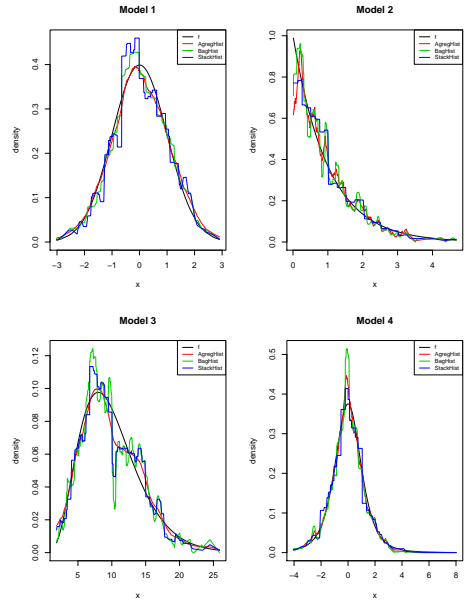}
\caption{Densities used in simulation models 1 to 4 together with the corresponding histogram and the estimators given by AggregHist, BagHist and StackHist. \label{dens1}}
\end{figure}

\begin{figure}[!ht]
\centering
\includegraphics[scale=0.32]{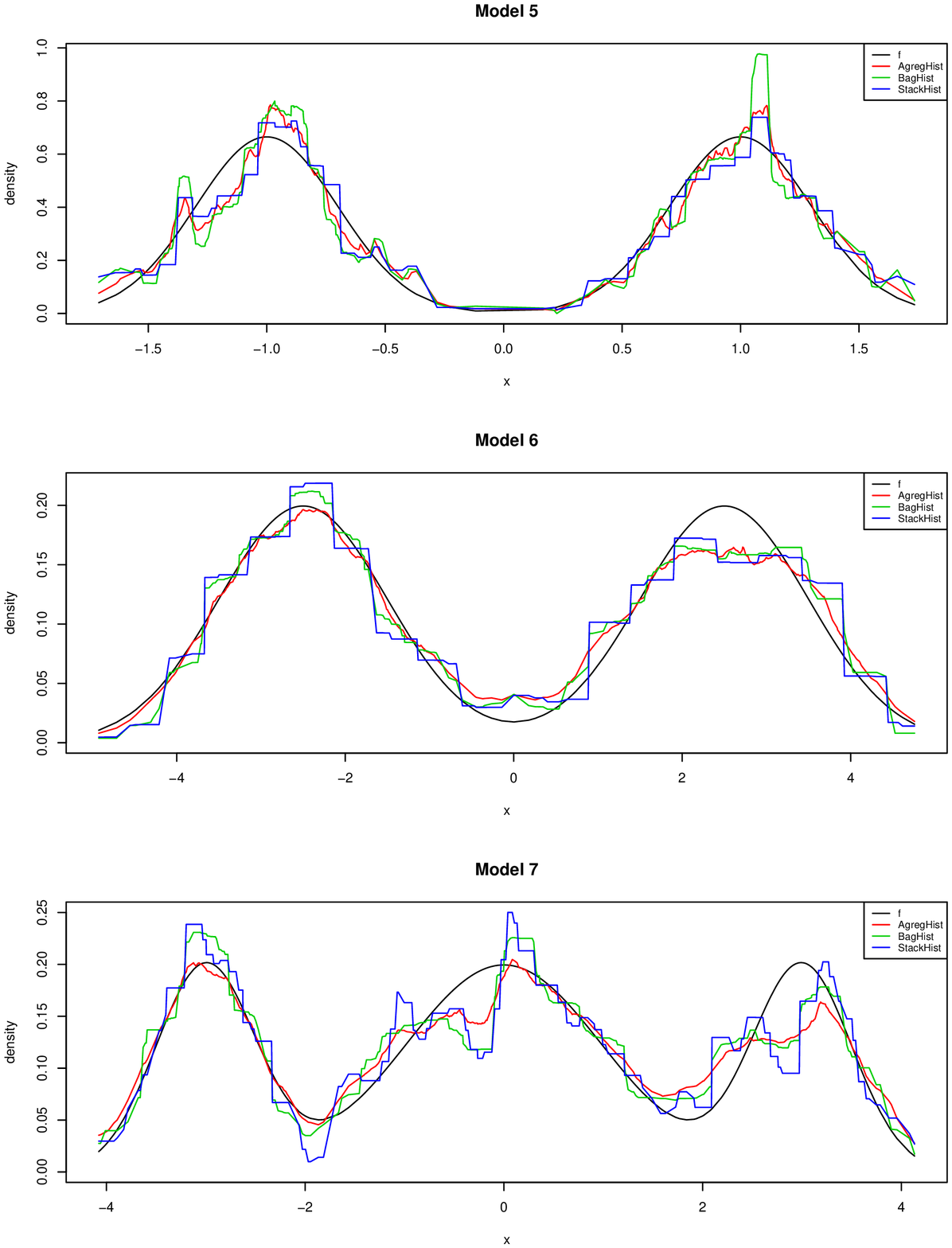}
\caption{Densities used in simulation models 5 to 7 together with the corresponding histogram and the estimators given by AggregHist, BagHist and StackHist.\label{dens2}}
\end{figure}

\begin{figure}[!ht]
\centering
\includegraphics[scale=0.4]{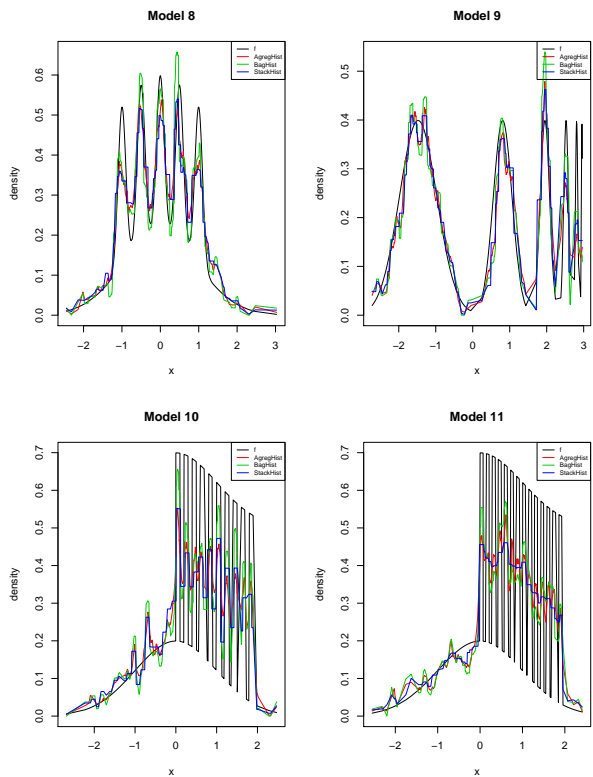}
\caption{Densities used in simulation models 8 to 11 together with the corresponding histogram and the estimators given by AggregHist, BagHist and StackHist. \label{dens3}}
\end{figure}

 {\it AggregHist} and {\it BagHist} give more smooth estimators than {\it StackHist}.

\newpage

\subsection{Tuning the algorithms}

For the existing algorithms we have used the same values suggested by their corresponding authors :
\begin{itemize}
\item For {\it Stacking}, six kernel density estimators are aggregated, three of them use Gaussian kernels with fixed bandwidths $h=0.1,0.2,0.3$ and three triangular kernels with bandwidths $h=0.1,0.2,0.3$. The number of cross validation samples is fixed to $V=10$.
\item For {\it AggPure} six kernel density estimators are aggregated having bandwidths $0.001,$ $0.005, 0.01, 0.05,0.1$ and $0.5$. We use the EM algorithm to optimize the coefficients of the linear combination. The final estimator is a mean over $S=10$ random splits of the original data set.
\item For {\it Boostkde}, we use $5$ steps for the algorithm aggregating kernel density estimators whose bandwidths are optimized using Silverman rule. Normalization of the output is done using numerical integration. Extensive simulations showed that more steps give rise to less accurate estimators.
\end{itemize}

Simple one kernel density estimators are also used in our comparisons using optimized bandwidths following Silverman rule ({\em KdeNrd0}) and unbiased cross validation ({\em KdeUCV}). \\
For the Histogram, fixed breaks are systematically used and their number is optimized over a fixed grid, retaining the one which maximizes the log likelihood of the obtained histogram over a test sample drawn from the same distribution as the learning sample. \\
The tuning parameters for our algorithms, the number of breakpoints and the value of $\gamma$ are optimized testing different values for each of them over a fixed grid. We test $10, 20$ and $50$ equally spaced breakpoints for each case. For $\gamma$ we chose the grid $0.5,1,1.5,2,2.5$. The best combination retained for each model is the one which maximizes the log likelihood over $100$ independent test samples drawn from the corresponding model. For $BagHist$ and $AggregHist$ we aggregate $M=300$ histograms. The optimal values for the histogram and for our algorithms are give in table \ref{valpars}. We denote the optimal number of breaks $L_H, L_{BH},L_{AH}$ for the Histogram, {\it BagHist} and {\it AggregHist} respectively, and $\gamma_{AH}$ the perturbation coefficient for {\it AggregHist}.

\begin{table}[!ht]
 \begin{center}
 \begin{tabular}{|r|rrrr|rrrr|rrrr|}\hline\hline
 \multicolumn{1}{|c|}{} &\multicolumn{4}{|c|}{\rule[-.3cm]{0cm}{.8cm} {\bf $n=100$}} & \multicolumn{4}{|c|}{\rule[-.3cm]{0cm}{.8cm} {$n=500$}} & \multicolumn{4}{|c|}{\rule[-.3cm]{0cm}{.8cm} {$n=1000$}}\\ \hline
\multicolumn{1}{c}{}&\multicolumn{1}{c}{$L_H$}&\multicolumn{1}{c}{$L_{AH}$}&\multicolumn{1}{c}{$L_{BH}$}&\multicolumn{1}{c}{$\gamma_{AH}$}&\multicolumn{1}{c}{$L_H$}&\multicolumn{1}{c}{$L_{AH}$}&\multicolumn{1}{c}{$L_{BH}$}&\multicolumn{1}{c}{$\gamma_{AH}$}
&\multicolumn{1}{c}{$L_H$}&\multicolumn{1}{c}{$L_{AH}$}&\multicolumn{1}{c}{$L_{BH}$}&\multicolumn{1}{c}{$\gamma_{AH}$}\\
\hline
$\mathcal{M}1$ & $50$ & $10$ &$50$&$1.0$&$50$&$10$&$10$&$0.5$&$50$&$20$&$20$&$0.5$\tabularnewline
$\mathcal{M}2$&$50$&$10$&$50$&$0.5$&$50$&$50$&$50$&$0.5$&$50$&$50$&$50$&$0.5$\tabularnewline
$\mathcal{M}3$&$50$&$10$&$50$&$0.5$&$50$&$10$&$50$&$0.5$&$50$&$20$&$50$&$0.5$\tabularnewline
$\mathcal{M}4$&$50$&$20$&$50$&$0.5$&$50$&$50$&$50$&$0.5$&$50$&$50$&$50$&$0.5$\tabularnewline
$\mathcal{M}5$&$50$&$20$&$50$&$1.0$&$50$&$50$&$50$&$2.0$&$50$&$50$&$50$&$1.0$\tabularnewline
$\mathcal{M}6$&$50$&$10$&$50$&$1.0$&$50$&$20$&$20$&$1.0$&$50$&$50$&$20$&$2.0$\tabularnewline
$\mathcal{M}7$&$50$&$20$&$10$&$0.5$&$20$&$20$&$20$&$0.5$&$20$&$50$&$20$&$0.5$\tabularnewline
$\mathcal{M}8$&$50$&$20$&$50$&$0.5$&$50$&$50$&$50$&$0.5$&$50$&$50$&$50$&$2.0$\tabularnewline
$\mathcal{M}9$&$50$&$20$&$50$&$0.5$&$50$&$50$&$50$&$0.5$&$50$&$50$&$50$&$1.0$\tabularnewline
$\mathcal{M}10$&$50$&$50$&$50$&$0.5$&$50$&$50$&$50$&$0.5$&$50$&$50$&$50$&$0.5$\tabularnewline
$\mathcal{M}11$&$50$&$20$&$50$&$0.5$&$50$&$50$&$50$&$0.5$&$50$&$50$&$50$&$0.5$\tabularnewline
\hline
\end{tabular}
\caption{Optimal parameters values used for our algorithms. \label{valpars}}
\end{center}
\end{table}

Finally, for $StackHist$ we aggregate six histograms having $5,10,20,30,40$ and $50$ equally spaced breakpoints. A ten fold cross validation is used.

\subsection{Results}

\begin{figure}[!ht]
\centering
\includegraphics[scale=0.35]{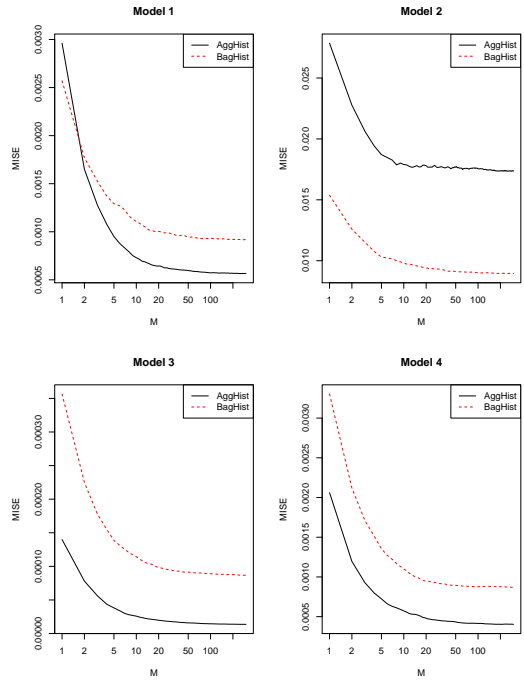}
\caption{MISE error versus number of aggregated histograms for Models 1 to 4. \label{evol1}}
\end{figure}

\begin{figure}[!ht]
\centering
\includegraphics[scale=0.35]{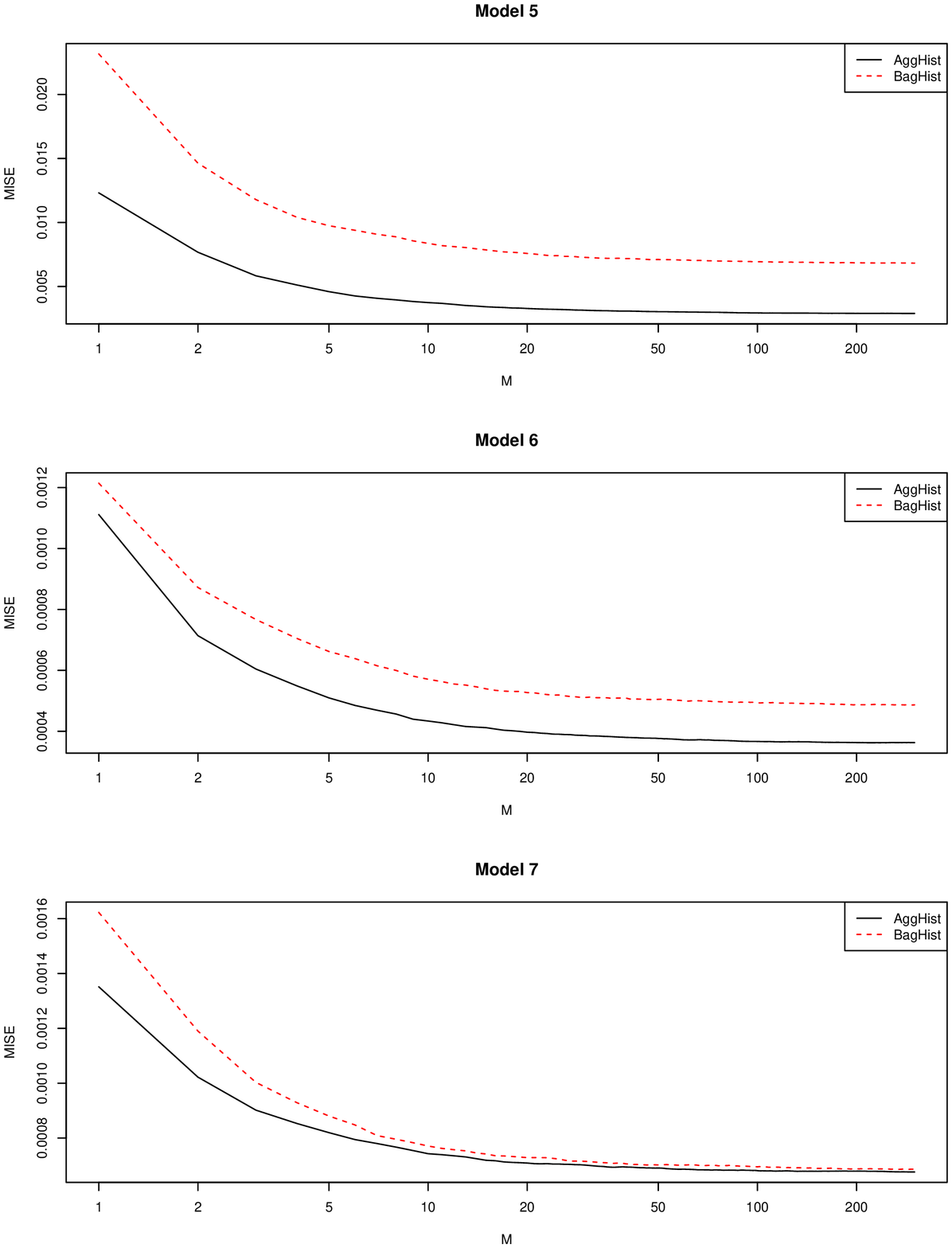}
\caption{MISE error versus number of aggregated histograms for Models 5 to 7. \label{evol2}}
\end{figure}

\begin{figure}[!ht]
\centering
\includegraphics[scale=0.35]{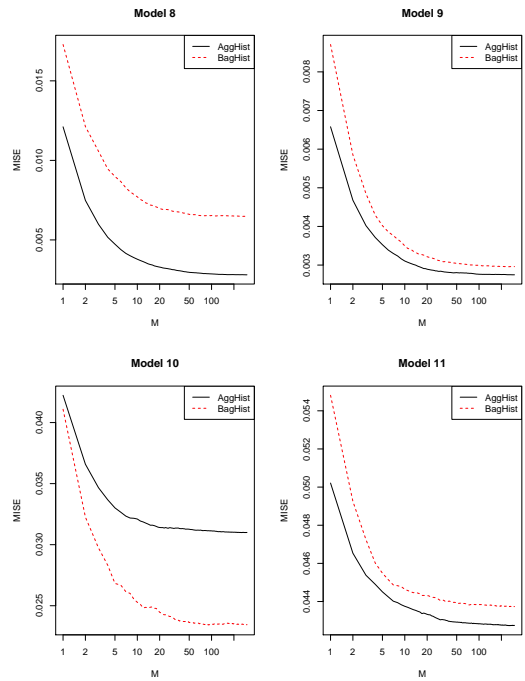}
\caption{MISE error versus number of aggregated histograms for Models 8 to 11. \label{evol3}}
\end{figure}

The performance of each model is evaluated using the Mean Integrated Squared Error (MISE). It is estimated as the average of the integrated squared error over $100$ simulations. First, for both {\it AggregHist} and {\it BagHist} we analyze the effect of the number $M$ of histograms aggregated. Figures \ref{evol1}, \ref{evol2} and \ref{evol3} show how the MISE varies when increasing the number of histograms. These graphics show clearly the contribution of the aggregation to the reduction of the MISE. For all the models, the error does not decrease significantly after about 100 iterations.

\newpage
\clearpage
We compare now our algorithms {\it BagHist}, {\it AggregHist} and {\it StachHist} to the following methods: the Histogram, {\it KdeNrd0}, {\it KdeUCV}, {\it Stacking}, {\it AggPure} and {\it BoostKde}. We have limited the comparisons to the ensemble methods which aggregated non parametric density estimators.\\

Tables 1 to 3 give the values of $100\times MISE$ for each method and simulation model for the three values of $n$. The best performances are indicated in bold.\\
In most cases, aggregation models have higher accuracy than simple methods like the histogram and KDE. However this is not the case for model $\mathcal{M}3$: KDE has a better performance, that is probably due to border effects. {\it BoostKde} gives in general better estimates for mixture models. All the methods have better accuracy in general when increasing the sample size $n$. {\it BagHist} and {\it AggregHist} always outperform the optimal Histogram and in general both have better accuracy than {\it StackHist}. For $n=100$ {\it BagHist} and {\it AggregHist} outperform the other algorithms over the complicated models ($\mathcal{M}8-\mathcal{M}10$). For $n=500$ and $n=1000$, {\it AggPure} outperforms the other methods for the last two models.
 Although our algorithms do not always outperform the other methods, their precision are not far from the best ones.\\

\input{synthese2}

\newpage

\section{Conclusion}
In this work we present three new algorithms for density estimation aggregating histograms. Two of them aggregate histograms over bootstrap samples of the data or randomly perturbed breakpoints. The third is a simple adaptation of the stacking algorithm where histograms are used instead of kernel density estimators. We have shown using extensive simulations that these algorithms and the other ensembles techniques have better accuracy than the histogram or KDE. The first two algorithms {\it BagHist} and {\it AggregHist} are very simple to implement, depend on very few parameters, and their computation complexity is proportional to that of a histogram. Theoretical properties of these algorithms are under study. Most of the algorithms described in this work may be easily generalized to the multivariate case.

\bibliography{AggregDens}

\bibliographystyle{plain}

\end{document}

%% file: synthese2.tex
\begin{table}[!ht]
 \small
 \caption{$n = 100 , M= 300, 100 \times$ MISE\label{tab100}}
 \begin{center}
 \begin{tabular}{lrrrrrrrrr}\hline\hline
100&{\scriptsize Histogram}&{\scriptsize KdeNrd0}&{\scriptsize KdeUCV}&{\scriptsize BoostKde}&{\scriptsize AggPure}&{\scriptsize Stacking}&{\scriptsize StackHist}&{\scriptsize BagHist}&{\scriptsize AggregHist}\tabularnewline
\hline
$\mathcal{M}1$&$2.870$&$\mathbf{0.193}$&$0.221$&$0.831$&$0.365$&$0.233$&$0.511$&$2.250$&$0.209$\tabularnewline
$\mathcal{M}2$&$4.820$&$7.040$&$4.710$&$9.320$&$2.750$&$2.610$&$1.580$&$4.330$&$\mathbf{1.560}$\tabularnewline
$\mathcal{M}3$&$0.162$&$\mathbf{0.012}$&$0.013$&$0.035$&$0.106$&$0.061$&$0.032$&$0.127$&$0.015$\tabularnewline
$\mathcal{M}4$&$1.380$&$\mathbf{0.181}$&$0.219$&$1.460$&$0.318$&$0.202$&$0.370$&$1.130$&$0.220$\tabularnewline
$\mathcal{M}5$&$6.700$&$6.410$&$3.070$&$\mathbf{0.387}$&$1.040$&$0.856$&$1.940$&$5.040$&$0.911$\tabularnewline
$\mathcal{M}6$&$0.739$&$0.149$&$0.082$&$\mathbf{0.056}$&$0.189$&$0.111$&$0.178$&$0.550$&$0.080$\tabularnewline
$\mathcal{M}7$&$0.820$&$0.278$&$0.156$&$\mathbf{0.107}$&$0.163$&$0.110$&$0.168$&$0.126$&$0.128$\tabularnewline
$\mathcal{M}8$&$4.310$&$2.280$&$2.130$&$3.840$&$1.910$&$1.620$&$1.850$&$3.390$&$\mathbf{1.560}$\tabularnewline
$\mathcal{M}9$&$2.460$&$2.330$&$1.350$&$1.350$&$0.967$&$0.927$&$1.130$&$1.950$&$\mathbf{0.832}$\tabularnewline
$\mathcal{M}10$&$5.340$&$6.750$&$6.940$&$6.090$&$4.480$&$4.700$&$4.780$&$\mathbf{3.640}$&$3.720$\tabularnewline
$\mathcal{M}11$&$6.130$&$6.620$&$6.840$&$5.590$&$5.160$&$5.650$&$5.590$&$\mathbf{4.940}$&$5.130$\tabularnewline
\hline
\end{tabular}

\end{center}

\end{table}

\begin{table}[!ht]
 \small
 \caption{$n = 500 , M= 300 , 100 \times$ MISE\label{tab500}}
 \begin{center}
 \begin{tabular}{lrrrrrrrrr}\hline\hline
500&{\scriptsize Histogram}&{\scriptsize KdeNrd0}&{\scriptsize KdeUCV}&{\scriptsize BoostKde}&{\scriptsize AggPure}&{\scriptsize Stacking}&{\scriptsize StackHist}&{\scriptsize BagHist}&{\scriptsize AggregHist}\tabularnewline
\hline
$\mathcal{M}1$&$0.474$&$0.085$&$0.100$&$0.098$&$0.076$&$\mathbf{0.054}$&$0.154$&$0.085$&$0.056$\tabularnewline
$\mathcal{M}2$&$0.786$&$5.980$&$2.840$&$6.930$&$1.280$&$1.120$&$\mathbf{0.545}$&$0.701$&$0.720$\tabularnewline
$\mathcal{M}3$&$0.024$&$\mathbf{0.003}$&$\mathbf{0.003}$&$0.029$&$0.020$&$0.012$&$0.009$&$0.018$&$0.004$\tabularnewline
$\mathcal{M}4$&$0.180$&$0.099$&$0.113$&$1.160$&$0.059$&$\mathbf{0.043}$&$0.114$&$0.131$&$0.072$\tabularnewline
$\mathcal{M}5$&$1.190$&$5.070$&$2.090$&$\mathbf{0.132}$&$0.430$&$0.254$&$0.591$&$0.870$&$0.424$\tabularnewline
$\mathcal{M}6$&$0.133$&$0.094$&$0.037$&$\mathbf{0.014}$&$0.036$&$0.024$&$0.056$&$0.040$&$0.024$\tabularnewline
$\mathcal{M}7$&$0.079$&$0.206$&$0.085$&$\mathbf{0.021}$&$0.036$&$0.035$&$0.067$&$0.050$&$0.034$\tabularnewline
$\mathcal{M}8$&$0.859$&$2.120$&$1.770$&$2.010$&$0.768$&$0.588$&$0.683$&$0.547$&$\mathbf{0.531}$\tabularnewline
$\mathcal{M}9$&$0.594$&$2.020$&$0.896$&$1.070$&$0.443$&$0.393$&$0.428$&$0.488$&$\mathbf{0.377}$\tabularnewline
$\mathcal{M}10$&$3.680$&$6.380$&$5.690$&$5.640$&$\mathbf{2.260}$&$2.740$&$3.790$&$2.800$&$3.420$\tabularnewline
$\mathcal{M}11$&$4.700$&$6.350$&$6.120$&$5.880$&$\mathbf{2.760}$&$4.430$&$4.900$&$4.160$&$4.000$\tabularnewline
\hline
\end{tabular}

\end{center}

\end{table}

\begin{table}[!ht]
 \small
 \caption{$n = 1000 , M= 300 , 100 \times$ MISE\label{tab1000}}
 \begin{center}
 \begin{tabular}{lrrrrrrrrr}\hline\hline
1000&{\scriptsize Histogram}&{\scriptsize KdeNrd0}&{\scriptsize KdeUCV}&{\scriptsize BoostKde}&{\scriptsize AggPure}&{\scriptsize Stacking}&{\scriptsize StackHist}&{\scriptsize BagHist}&{\scriptsize AggregHist}\tabularnewline
\hline
$\mathcal{M}1$&$0.207$&$0.066$&$0.081$&$0.136$&$0.039$&$\mathbf{0.032}$&$0.087$&$0.064$&$0.036$\tabularnewline
$\mathcal{M}2$&$0.374$&$5.690$&$2.500$&$6.090$&$0.896$&$0.850$&$0.365$&$\mathbf{0.339}$&$0.676$\tabularnewline
$\mathcal{M}3$&$0.011$&$\mathbf{0.002}$&$\mathbf{0.002}$&$0.014$&$0.009$&$0.006$&$0.005$&$0.008$&$\mathbf{0.002}$\tabularnewline
$\mathcal{M}4$&$0.099$&$0.071$&$0.076$&$0.816$&$0.036$&$\mathbf{0.028}$&$0.074$&$0.062$&$0.035$\tabularnewline
$\mathcal{M}5$&$0.593$&$4.530$&$1.730$&$\mathbf{0.062}$&$0.258$&$0.146$&$0.318$&$0.420$&$0.216$\tabularnewline
$\mathcal{M}6$&$0.064$&$0.078$&$0.030$&$\mathbf{0.008}$&$0.018$&$0.013$&$0.032$&$0.023$&$0.024$\tabularnewline
$\mathcal{M}7$&$0.054$&$0.181$&$0.066$&$\mathbf{0.012}$&$0.021$&$0.020$&$0.042$&$0.031$&$0.031$\tabularnewline
$\mathcal{M}8$&$0.590$&$2.070$&$1.690$&$1.530$&$0.498$&$0.381$&$0.506$&$\mathbf{0.336}$&$0.506$\tabularnewline
$\mathcal{M}9$&$0.406$&$1.890$&$0.767$&$0.902$&$0.307$&$\mathbf{0.286}$&$0.331$&$0.332$&$0.342$\tabularnewline
$\mathcal{M}10$&$3.830$&$6.220$&$5.170$&$5.360$&$\mathbf{1.600}$&$2.450$&$3.880$&$2.980$&$3.490$\tabularnewline
$\mathcal{M}11$&$4.700$&$6.220$&$5.590$&$5.340$&$\mathbf{1.950}$&$4.100$&$4.790$&$4.380$&$4.070$\tabularnewline
\hline
\end{tabular}

\end{center}

\end{table}

%% file: AggregDens2012-07-17Arxiv.bbl
\begin{thebibliography}{10}

\bibitem{Brei-bag}
L.~Breiman.
\newblock Bagging predictors.
\newblock {\em Machine Learning}, 24(2):123--140, 1996.

\bibitem{Brei-stack}
L.~Breiman.
\newblock Stacked regression.
\newblock {\em Machine Learning}, 24(1):49--64, 1996.

\bibitem{RF}
L.~Breiman.
\newblock Random forests.
\newblock {\em Machine Learning}, 45(1):5--32, October 2001.

\bibitem{DiMarz}
M.~Di~Marzio and C.C. Taylor.
\newblock {Boosting kernel density estimates: A bias reduction technique?}
\newblock {\em Biometrika}, 91(1):226--233, 2004.

\bibitem{Freund}
Y.~Freund.
\newblock Boosting a weak learning algorithm by majority.
\newblock {\em Information and Computation}, 121,2:256--285, 1995.

\bibitem{Jones}
M.~Jones, O.~Linton, and J.~Nielsen.
\newblock A simple bias reduction method for density estimation.
\newblock {\em Biometrika}, 82:327--338, 1995.

\bibitem{Marron}
J.S Marron and M.P. Wand.
\newblock Exact mean integrated square error.
\newblock {\em The Annals of Statistics}, 20(2):712--736, 1992.

\bibitem{Ridgeway}
G.~Ridgeway.
\newblock Looking for lumps: boosting and bagging for density estimation.
\newblock {\em Comput. Stat. Data Anal.}, 38(4):379--392, 2002.

\bibitem{Rigo}
P.~Rigollet and A.~B. Tsybakov.
\newblock Linear and convex aggregation of density estimators.
\newblock {\em Math. Methods Statist.}, 16(3):260--280, 2007.

\bibitem{Rosset}
S.~Rosset and E.~Segal.
\newblock Boosting density estimation.
\newblock In {\em In Advances in Neural Information Processing Systems 15},
  pages 641--648. MIT Press, 2002.

\bibitem{Smyth}
P.~Smyth and D.~Wolpert.
\newblock Linearly combining density estimators via stacking.
\newblock {\em Mach. Learn.}, 36(1-2):59--83, 1999.

\bibitem{Song}
X.~Song, K.~Yang, and M.~Pavel.
\newblock Density boosting for gaussian mixtures.
\newblock {\em Neural Information Processing}, 3316:508--515, 2004.

\bibitem{Wolpert}
D.H. Wolpert.
\newblock {Stacked Generalization}.
\newblock {\em Neural Networks}, 5:241--259, 1992.

\end{thebibliography}
